\documentclass{aastex62}

\usepackage[T1]{fontenc}
\usepackage{amsmath}

\graphicspath{{./}{figures/}}

\received{October 19, 2019}
\revised{February 12, 2020}
\accepted{February 18, 2020}
\submitjournal{ApJ}

\shorttitle{Equatorial scattering region measurement}
\shortauthors{Shablovinskaya et al.}

\begin{document}

\title{Measuring the AGN sublimation radius with a new approach: \\ reverberation mapping
of the broad line polarization}

\correspondingauthor{Elena S. Shablovinskaya}
\email{e.shablie@yandex.com}

\author{Elena S. Shablovinskaya}
\affil{Special Astrophysical Observatory of the Russian Academy of Sciences \\ Nizhnii Arkhyz, 369167, Russia}

\author{Viktor L. Afanasiev}
\affil{Special Astrophysical Observatory of the Russian Academy of Sciences \\ Nizhnii Arkhyz, 369167, Russia}

\author{Luka \v C.  Popovi\'c}
\affiliation{Astronomical Observatory \\ Volgina 7, 11060 Belgrade 74, Serbia}

\begin{abstract}

Here we give an observational method for measurements of {the equatorial scattering region} radius using variability in the polarized broad lines in Type 1 active galactic nuclei (AGNs). The polarization in broad lines of Type 1 AGNs is mostly caused by equatorial scattering, where specific features allow one to separate its contribution from the total polarized flux. We propose to monitor variability in the polarized line flux and find the time lag between the nonpolarized continuum and polarized broad line variability. The distance to the scattering screen can then be determined from the time delay.

The method was, for the first time, applied to the observations of the Type 1 AGN Mrk 6, and we found that the size of the scattering region in this AGN is around 100 light days. That is significantly smaller than the dusty region size estimated by the infrared interferometric observations and also larger than known broad line region (BLR) size. This indicates that the scattering region lies between the BLR and the dusty region and could be used as a probe of the dust sublimation radius.

\end{abstract}

\keywords{techniques: polarimetric -- galaxies: Seyfert}

\section{Introduction}

{In the 1970s the  Seyfert galaxy (Sy) class was divided into two subclasses according to the presence the absence of the broad permitted emission lines in their spectra. However, the discovery of the broad line component in the polarized spectrum of the archetypal Sy 2 NGC 1068 \citep{AntMil85} established of the common model of active galactic nuclei (AGNs). Within this "unified model"\ \citep{Ant93, UnMod}, the dichotomy of the observed AGN properties was explained by the orientation of the nuclei relative to the observer, and the inner structure was postulated to be generally the same for different AGN types. In the central AGN parts the fast rotating gas clouds radiate the emission lines (the broad line region, or BLR) and are surrounded by external obscuring matter, a so-called "dusty torus"\, predicted earlier by several authors \citep{RR77,Keel80}. Currently, the "unified model"\ is mostly accepted, yet there are peculiar types of AGNs whose properties are not clarified by specific orientation e.g. narrow-line Sy 1 \citep{NL1, NL2} and changing-look AGNs \citep{CL}. Later, the existence of the dusty circumnuclear region was approved with infrared (IR) observations \citep[see][and references therein]{grav19}; also, it was discovered that this geometrically and optically thick region has a more complicated form and structure than the "torus" suggested earlier \citep{honig19}.}

{Central AGN parts remain optically unresolved. As a result, the region where the BLR transits into the dusty region is comparatively poorly studied. The lack of the data on the temperature, chemical composition, and the distances inside the central regions limits the usage of the theoretical models \citep[e.g.][]{bar87, Bar92} to locate the dust sublimation radius in an AGN structure. On the contrary, in the most cases the theoretical predictions are not in an agreement with the observational data. }

{\citet{Smith04} and \citet{GG07} proposed that the flattened disk-like region is lying in the equatorial plane between the "dusty torus"\ and the BLR. The region is assumed to consist of free electrons and a small percentage of dust grains provided for the explanation of the preferable polarization vector direction (tending to be parallel to the radio axis) in Type 1 AGNs\footnote{orientated so that the inner parts of the nucleus are available for the observer.}. This so-called equatorial scattering region is also responsible for the specific polarization signatures along the emission line profiles: $S$-shaped swing in the polarization angle and a dip in the polarization degree along the emission line profile \citep{GM94}, while without the scattering emission, the line should be polarized in the same way as the nearby continuum. As the angle swing is connected with the rotation of the gas clouds  in the BLR, \citet{AfPo14} demonstrated that based on the circular velocities of the gas in the BLR obtained by spectropolarimetric measurements of the emission line, the mass of the central supermassive black hole (SMBH) could be estimated, and this estimate would not be strongly influenced by the inclination angle \citep[see][]{sa18}. }

{The existence of the scattering region in the equatorial plane in central AGN parts is one more piece of evidence of the smooth transition of the BLR to the dusty region with a gradual increase of the optical depth $\tau$ (the modeling shows that in the scattering region $\tau$ should be > 0.1 \citep{GG07} and 1<$\tau$<3 \citep{Marin12}). However, now the distance to the equatorial scattering region $R_{\rm sc}$ is not well known -- usually $R_{\rm sc}$ is assumed to be equal to the size  of the inner part of the "dusty torus" obtained by IR observations (see Sec. \ref{sDT}).}

{The aim of this work is to explore the possibility of finding the size of the scattering (dust sublimation) region in AGNs. In this work, we propose and demonstrate a new technique for the measurements of the equatorial scattering region radius that uses reverberation mapping of the broad emission lines in the polarized light. The paper is organized as follows. In Sec. \ref{ladder} we describe the methods for finding the distances inside the AGN, while in Sec. \ref{main} the new method of size estimation of the scattering region  is introduced. Sec. \ref{test} is dedicated to the observational test of our method for the case of the AGN Mrk 6. In Sec. \ref{disc} a critical discussion of the method suggested in the paper is given within an analysis of the previous studies in the same field, and Sec. \ref{conc} provides a summary of the main result. }

\section{Sizes in AGN\lowercase{s}}
\label{ladder}
\subsection{Measurement of the BLR size}
\label{sBLR}
{Central regions of AGNs remain optically unresolved, therefore the dimensions of the BLR is not well known. However, there is currently an understanding of the size ladder inside an AGN.}

{This important step was done with the help of the reverberation mapping. It was found that the central continuum source is surrounded by the BLR, and the response of the emission line radiation to the continuum variability is delayed in a time interval $\Delta t$ that characterizes the BLR size: $R_{\rm BLR} = c \Delta t$ \citep{ChL,BlandMcKee93,Pet93}. Soon after, the dependency of $R_{\rm BLR}$ on the ionizing source luminosity was discovered \citep{KorGas91}. Then, \citet{NetPet97}, using data from \citet{Wan97}, found that the time delay is longer when the ionization potential is lower, which corresponds to the BLR stratification. Later, this was statistically strengthened with the reverberation monitoring campaigns \citep[see e.g.][]{Bentz13}. The signal delay in the H$\alpha$ line exceeds the delay observed in other broad lines \citep[it could be found by the comparison of the observational data of more than 40 galaxies, see][]{cat} and, therefore, characterizes, more or less, the BLR border. However, using the reverberation mapping method one measures the time lag corresponding to the maximum of the cross-correlation function. 
Therefore, the sizes measured from the delay between the continuum and line fluxes, the so-called photometric BLR radius \citep[see][]{KorGas91, Kish09}, might give an underestimated $R_{\rm BLR}$ (e.g., due to the BLR inclination).} 

\subsection{Measurement of the dusty region size}
\label{sDT}

{As the distance from the central source increases while the energy decreases, free electrons are recombined, neutral gas is dominant, and the dust grains appear while the temperature drops down. The dusty circumnuclear region (or the "dusty torus") is observed in AGNs in the IR band. Currently, there are two methods to determine the radius of the inner dusty region. The first is the reverberation mapping that uses the response of the near-IR (usually $K$-band) signal to the ionizing (optical) continuum variability. Reverberation mapping observations are time consuming, additionally the dusty region size is a few times larger than the BLR, which means one needs more time to conduct the monitoring observations. For instance, the dusty region is four times larger than the BLR in the case of NGC 4051 \citep{Suganuma06,Den09,Zu11}, and five times larger for NGC 4151 \citep{Kish09,Pet04,maoz91}. Therefore, presently the IR reverberation mapping has currently succeeded for the measurment of the inner dusty region radii of $\sim$20 AGNs \citep{Minezaki04,Suganuma06,Koshida09,Koshida14, Mandal18}. The second technique is the near-IR and mid-IR interferometry resolution the central parts of an AGN (up to a few tenths of a parsec). This method was also used to on $\sim$24 AGNs \citep[][etc.]{Kish09,Kish11,grav19,wei12}. }

{\citet{Kish09} found the systematic difference between the size of the dusty region obtained with the different observational techniques. They suggested that the reverberation mapping reveals the parts of the dusty region closer to the central source than the region of the maximum brightness temperature in the $K$ band determined by the interferometry (especially using the thin-ring model in the case of the extended region). Furthermore, \citet{Kish09} called the result obtained by the Keck interferometer as a "probing the dust sublimation radius". In fact, the IR radiation of AGNs is well explained with the thermal dust emission, and both of the IR methods are dedicated to the observations of the region where the dust is likely to be dominant, and remote to the size of $R_{\rm dust}$ from the central engine. So, it seems that this radius represents the upper limit of the dust sublimation radius. }

\subsection{Scattering region and polarization in Sy 1 galaxies}
\label{sPOL}

\begin{figure}
    \centering
    \includegraphics[width=0.9\textwidth]{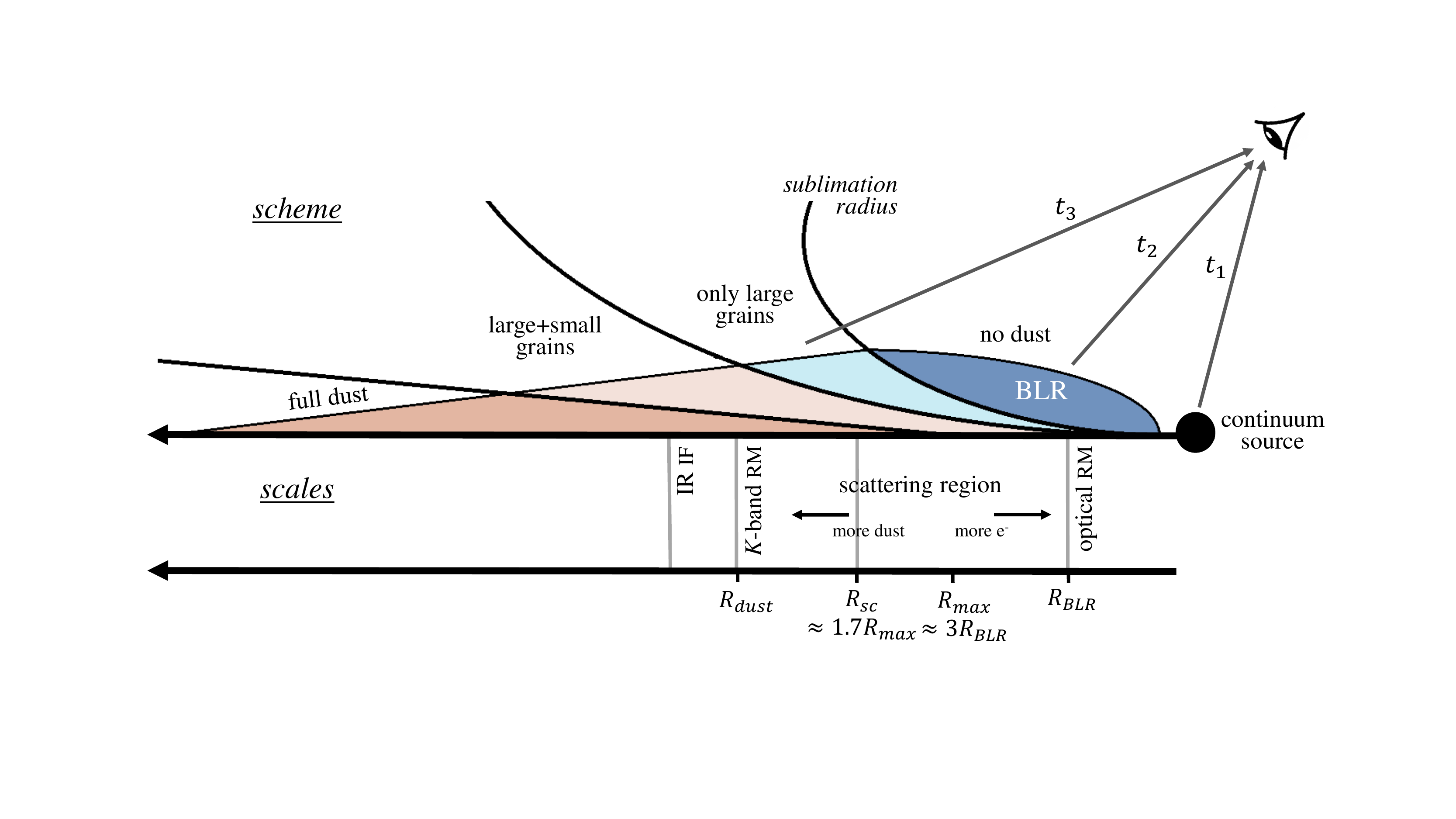}
    \caption{{The scheme of the inner AGN sizes. The upper part of the figure illustrates the scheme of the dust distribution inside the AGN introduced by \citet{dust}. $t_1$, $t_2$, and $t_3$ correspond to the moment when a signal comes from the central ionizing source, the BLR, and the scattering region, respectively. The lower part marks the approximate size scale connecting the photometric BLR radius $R_{\rm BLR}$ observed with the reverberation mapping technique ("optical RM"), the maximal radius of the BLR $R_{\rm max}$, the size of the scattering region, and the radius of the inner part of the dusty region $R_{\rm dust}$ measured by the IR reverberation mapping technique ("$K$-band RM") and by the near-IR interferometry ("IR IF"). The scale relations are from \citet{APS19}.  }}
    \label{fig1}
\end{figure}

{The presence of the equatorial scattering region, on the one hand, and the theoretical model of the BLR extension up to the dusty region \citep[see][and references therein]{Netzer15}, on the other, points out that in going farther from the ionizing source the free electron domination changes smoothly into the dust domination; $
\tau$ also increases gradually which leads to a scattering "screen"\ with blurred borders. The radial extension and the location of the "screen"\ are not determined, but from the previous studies they should lie within the limits $R_{\rm BLR} \leq R_{\rm sc} \leq R_{\rm dust}$. The existing models \citep{Marin12, sa18} are based on the assumption that the Thomson scattering dominates in the equatorial scattering region, yet the contribution of the Mie and Rayleigh scattering is also taken into account. According to the dominant scattering type, one could conclude that $R_{\rm sc}$ corresponds to the lower limit of the dust sublimation radius. }

{ As the equatorial scattering region was introduced to explain the origin of the specific features in polarization in Type 1 AGNs, the idea to apply the reverberation mapping method to the polarized flux soon appeared. The modeling performed by \citet{Goos08} shows that the cross-correlation analysis should reveal the time lag between the total flux and one polarized due to the equatorial scattering, and that this lag should characterize the distance between the regions producing these fluxes\footnote{Note here that originally \citet{Goos08} were not careful about the wavelength band, and there was no word about whether the line or continuum emission should be observed.}. This idea was taken to the observations of Sy 1 NGC 4151 \citep{Gas12}, which had demonstrated the equatorial scattering features earlier \cite[see][]{Smith05}. The observations were conducted in $U$ total flux corresponding to the ionizing source and $B$ polarized flux (covering the continuum and broad H$\beta$ line). The cross-correlation analysis discovered the 8 days delay between the $U$-band total and $B$-band polarized flux. This result appeared to be comparable to the estimation of the BLR size in the H$\alpha$ or H$\beta$ line which is equal to 2-10 lt-days in NGC 4151 \citep{Pet04,maoz91,Sh08}. Moreover, the observed time lag was also noticeably less than the expected size of the inner border of the dusty region. This indicates that the scattering region is located near (or even in) the BLR and is consistent with the modern conception about inflows/outflows, a hot corona, and etc. in the AGN central regions. }

{To analyze this result, the mechanisms producing the polarization in Type 1 AGNs should be examined. There are two types of them that one could consider: "internal"\ and "external". The "internal"\ mechanisms are those that are located inside the dusty region. Among them there are the following: the radiative transfer in a magnetized accretion disk produces the polarization parallel to the disk rotational axis \citep[see, e.g.][]{Gnedin}; the synchrotron radiation of the jet that provides variable polarization when the polarization vector rotates with respect to the disk rotational axis; and the clouds of matter (e.g. of free electrons) that cause the polarization due to the several types of scattering (e.g. Compton, Thomson),  etc. The "external"\ mechanisms of polarization include only two processes: the equatorial (at the dusty "torus") and polar (at the ionization cone) scattering. Both of them could be observed together in one object, yet it is expected that equatorial scattering dominates in Type 1  and polar in Type 2 AGNs. Within this paper we consider only Type 1 AGNs (with broad lines) and therefore we expect to have dominant equatorial scattering (i.e. that the polar scattering is negligible). The contributions of all the mechanisms mentioned above are superposed and majorly could not be unambiguously divided. The same is true for the observations in broad bands (e.g. $B$ band as for \citet{Gas12}): the  multiple sources of polarization could make a contribution to the final result, and the equatorial scattering in the H$\beta$ broad line was only one of them. Since the calculated size was more than twice as small as than the BLR one, we could conclude that it could be the size of the scattering region but not a size of the \textit{equatorial} scattering region.}

{Additionally, it is important to raise a question here about the intrinsic (due to radiation transfer) polarization of broad line emissions. Unlike the magnetized accretion disks or relativistic jets, there are no obvious and strong polarization mechanisms inside the BLR. The BLR is considered to be a flattened structure consisting of gas clouds that are radiating independently, therefore there is not a dominant classical radiative transfer process and the BLR intrinsic polarization can be neglected. This assumption will be used further. However, let us consider several polarization mechanisms which may affect the broad line emission polarization. 
\begin{itemize}
    \item \textit{Transmission through a medium.} The polarization is sensitive to the  anisotropy of a medium. So, considering the BLR gas clouds' isotropy and the magnetic field absence, one can conclude the BLR emission is nonpolarized. It is important to note the limb polarization effect could take place (the Chandrasekhar-Sobolev effect) but it would vanish due to the integration of the emission from the whole region. 
    \item \textit{Scattering on the free electrons in the BLR..} The scattering at the free electrons inside the BLR could also appear. The photons coming from the BLR could be randomly scattered. However, (i) the random scattering events will be summed up within the observations and the total polarization produced by them will converge to zero; (ii) the Monte Carlo simulations show that the majority of photons are scattered only once at the equatorial scattering region \citep{sa18}. 
    \item \textit{Inflows/outflows.} \citet{sa18} have modeled the broad line polarization profile concerning inflows/outflows. They conclude that inflows/outflows have a negligible influence except in the extreme cases  where outflow is of an order of 1000 km/s, but in this case one should detect a  broad line strongly shifted to the blue. In the most of AGNs this is not the case.  
\end{itemize}}

{All these arguments allow us to consider within this work that the BLR broad lines are intrinsically  nonpolarized before the equatorial scattering. In contrast, if any of the intrinsic polarization mechanisms inside the BLR are by incident stronger and more important than assumed, then the BLR polarization should be constant and weaker than the one formed at the equatorial region (otherwise the equatorial scattering is not detectable for the source). In that case, the reverberation mapping may only give less contrast in the correlation peak, but the time lag between the continuum and the polarized line should be mostly driven by the external (equatorial) scattering. }

\section{New approach for the size determination of the scattering region}
\label{main}

The basic idea for the estimation of the equatorial scattering region's size is to measure the time lag between the continuum nonpolarized emission  and polarized broad emission line flux. Let us assume that the variable ionization continuum flux comes to the observer at time $t_1$ (see Fig. \ref{fig1}). The variability in the ionized continuum causes the broad emission line variability that an observer can detect at $t_2$. Then, since the emission line polarized light is scattered in the equatorial scattering region, the observer detects the change in the  polarized broad line flux at $t_3$. Therefore, the time lag $R_{\rm BLR}=c\Delta\tau_{12}=c\cdot(t_2-t_1)$ ($c$ is speed of light) gives the BLR photometric size, and $R_{\rm sc}=c\Delta\tau_{13}=c\cdot(t_3-t_1)$  gives the distance of the broad line scattering screen from the central continuum source, assumed as point-like. 

To use this method, the spectropolarimetric observations are preferably required (we show an example in the following section). However{,} the spectropolarimetric observations are time consuming and they are technically demanding. Therefore, we propose the polarization photometric observations with narrow-band filters, which cover the line emission and nearby continuum. In any case, one should obtain a set of three Stokes parameters of {the} linear polarization for the spectral bands corresponding to the continuum and the emission line: $\{ I, Q, U\}_{\rm cont}$ and $\{ I, Q, U\}_{\rm line}$. To find the time delay needed for light to pass from the BLR to the scattering region, the nonpolarized continuum flux and polarized line flux should be compared. The nonpolarized continuum flux $I_{\rm cont}$ is equal to the total flux in all {observed} polarization directions (depending on the analyzer type) in the continuum band. As for the emission line flux, $I_{\rm line}$, the total flux could be found in the same way. To calculate the polarized line flux, $I_{\rm pol-line}$, use the following:

\begin{equation} \label{eu}
    I_{\rm pol-line} = I_{\rm line} \cdot P, \\ \ \ P=\sqrt{({Q}_{\rm line}-{Q}_{\rm cont})^2+ ({U}_{\rm line}-{U}_{\rm cont})^2},
\end{equation}
where {$P$ is the polarization degree, $0<P<1$.} Subtracting {the} Stokes parameters of the nearby continuum from the broad line Stokes parameters, one excludes the contribution of different polarization mechanisms to the polarization in broad line (as, e.g., instrumental, interstellar, etc.). {In case when the polarization in the broad line is not due to the equatorial scattering, all emissions of the central parts of the AGN (accretion disk, BLR, etc.) should be polarized in the same way, without any signatures of the polarization degree "bump"\ along the line profile, and according to the equation \ref{eu} the method will not show any variability in the polarized line flux. }

In following section we give an example of $R_\textrm{sc}$ measurements using spectropolarometric observations for the Mrk 6 AGN.

\section{Observational test}
\label{test}

{Mrk 6 ($z=0.019$) is }a well-known S0 galaxy with an active nucleus of the intermediate Sy 1.5 type \citep{OK76}. {In order to check the method availability we have used the spectropolarimetric data of Mrk 6 obtained within the 2010-2013 monitoring campaign \citep{Af14}.}

The observations and data reduction are described in detail in \citet{Af14}, and here we only give a short summary. The observational data used were obtained using the 6 m BTA telescope (SAO RAS, Russia) with the SCORPIO device \citep{AfMois05} in polarimetric and spectropolarimetric modes. {In spectropolarimetric mode, the Wollaston prism and the $\lambda/2$ phase plate were used with gratings VPHG1200 (spectral coverage 3700-7300\AA) and VPHG940 (spectral coverage 3700-8400\AA). In the polarimetric mode, the polaroid was used with the filter $V$ ($\lambda_{max}=5500$\AA, FWHM=850\AA).} Together with the object, the field stars were observed to measure the interstellar polarization and polarimetric standards were observed to control the polarization position angle. In each epoch a series of the images was shot to take into account the variation of the atmosphere. The method of polarimetric observations with SCORPIO is described in  \cite{AfAmir12}.

\begin{figure}
    \centering
    \includegraphics[scale=0.55, angle=90]{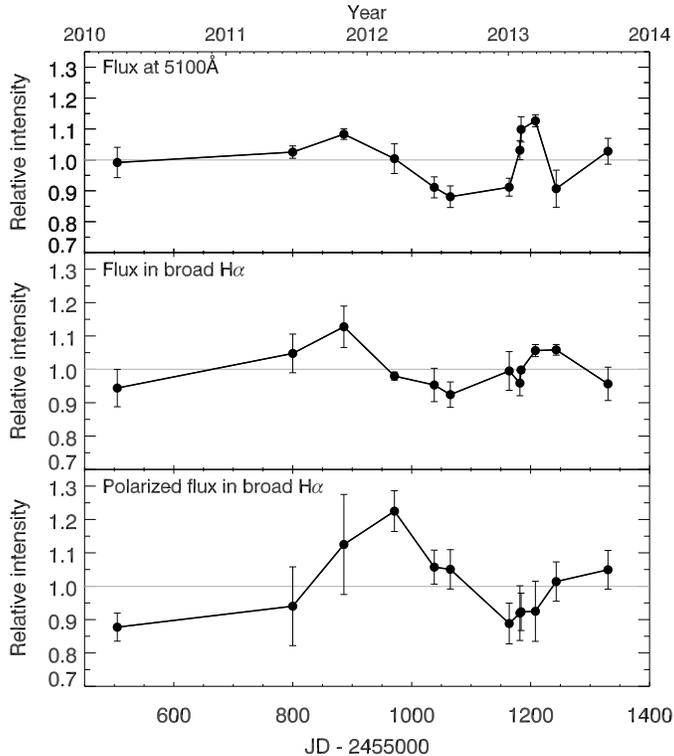}
    \caption{The light curves of Sy 1.5 Mrk 6: the total flux in continuum at 5100\AA\ (upper panel),  the total flux in the H${\alpha}$ emission line (middle panel), and  the polarized flux in the H${\alpha}$ emission line (bottom panel). }
    \label{fig2}
\end{figure}

We obtained Mrk 6 spectra from 12 epochs in the  2010-2013 monitoring period.  In Fig. \ref{fig2} we show (from the top to bottom) the light curves of the nonpolarized flux at $\lambda$5100\AA, corresponding to the ionizing continuum \citep{wo18}, the nonpolarized H$\alpha$ flux, and polarized H$\alpha$ flux, respectively.{ All light curves are shown with subtracted long-term trends and normalized, then the given units are in relative intensities. }

\begin{figure}
    \centering
    \includegraphics[angle=90,scale=0.45]{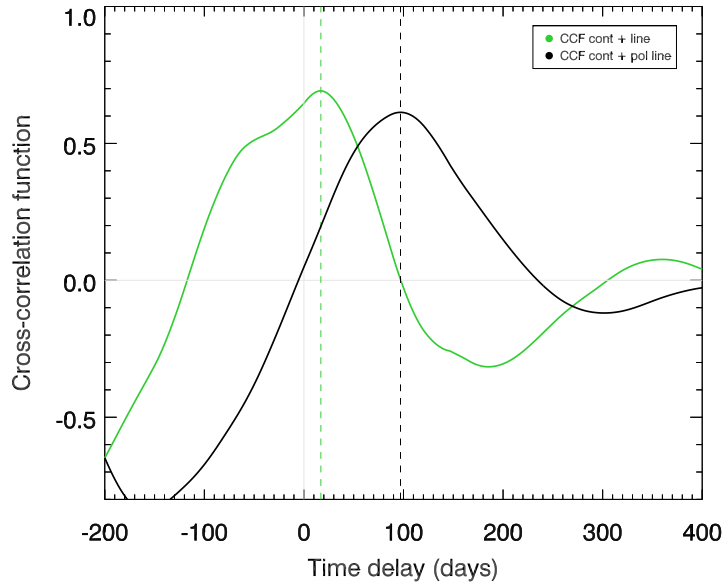}
    \caption{CCFs between the nonpolarized continuum flux and nonpolarized line flux (green line), and between the nonpolarized continuum flux and {polarized} broad line flux (black line).}
    \label{fig3}
\end{figure}

To estimate the time delays between the nonpolarized continuum flux and the polarized emission line flux, we used the cross-correlation analysis \citep{CCF}. Let us assume $F_\textrm{cont}$ is the nonpolarized continuum flux and $F_\textrm{pol-line}$ is the polarized emission line flux, and that both are finite discrete functions of length $N$. Then, the cross-correlation function (CCF) is calculated as:
$$
(\tilde{I}_\textrm{cont} \otimes \tilde{I}_\textrm{pol-line})[n] = \frac{1}{N} \sum_{j=0}^{N-1} \{\tilde{I}_\textrm{cont}\}_j \cdot \{\tilde{I}_\textrm{pol-line}\}_{j+n},
$$
where $\tilde{I}$ is the flux normalized at the dispersion and $\otimes$ corresponds to the convolution process. The position of the CCF maximum therefore corresponds to the time lag between the two functions.
The result of the cross-correlation analysis is shown in Fig. \ref{fig3}.
The black curve shows the cross-correlation function obtained for the nonpolarized continuum and {polarized} broad line flux. The time delay is equal to $\tau_{13}=94.7 \pm 6.9$ lt-days. Our result was also compared with the results given by {\tt JAVELIN} \citep{jav}: $\tau_{\rm JAV} = 99.5 \pm 3.6$ days, which is also in a good agreement. This time delay corresponds to the time required for the total continuum radiation formed in the central regions to reach the region of equatorial scattering -- the dusty torus. Also, by the same method, we obtained the time delay between the nonpolarized continuum and the total broad line flux -- i.e., by the classic BLR reverberation mapping method, and the estimation is equal to $\tau_{12}=13.0\pm6.4$ lt-days. This estimation is less that previously calculated in \citet{Af14} of $20.3^{+2.3}_{-2.1}$ lt-days, but it is still within the margin of error.  

{The error of the time delay $\tau_{13}$ obtained by the CCF was estimated by the bootstrapping method \citep{boot}. Bootstrapping is a statistical method based on random sampling with replacement. Each unit of the sample is assumed to be distributed within the confidence interval of the size of the flux error in the light curves (both the nonpolarized continuum and polarized broad line). Therefore, the flux determination uncertainty  is statistically considered. Synthesizing series of the investigating sample, one can estimate its statistical properties. In this work, the time series of the flux variations (in the continuum and the polarized line) are modeled by means of the bootstrapping and cross-correlated for a few thousand times. The CCF peak distribution converges to be a Gaussian one while increasing the number of modeling series. The maximum of this distribution is chosen as the most reliable time lag ($\tau_{13}=94.7$ lt-days). The FWHM of the Gaussian fit to the distribution function is assumed to be the error of the time lag $\delta \tau_{13}=6.9$ lt-days. In the case where the error is calculated as a 10\% level of the Gaussian fit, it is only twice as large as $\Delta \tau_{13} \approx 15 $ lt-days, and it is still much smaller than the suggested time lag. It demonstrates that the obtained result is essentially unaffected by the measurement errors.   }

\section{Discussion}
\label{disc}

Let us compare the obtained time delay with the previously calculated estimates of the size of the BLR and dust sublimation region in the galaxy Mrk 6. 

{\textit{a) Comparison with $R_{\rm BLR}$.} The size of the BLR region was found previously by the reverberation mapping in the broad line -- $R_{\rm BLR}=20.6 \pm 2.0$ lt-days \citep{Bentz13}. The estimates of the size of the BLR obtained in this paper, $\tau_{12}=13.0\pm6.4$lt-days and in \citet{Af14} for the same data, $20.3^{+2.3}_{-2.1}$  lt-days, are consistent with \citet{Bentz13}. Yet, note here that values of $R_{\rm BLR}$ are distributed, and the reasons could be both internal (the BLR size changes due to the nucleus activity) and "external"\ (not ideal cross-correlation methods, cadence, etc.).  }

{One can assume (and it is consistent with Fig. \ref{fig1}) that the maximal size of the BLR as a disk-like structure $R_{\rm max}$ is greater than the photometric BLR size (estimated by the photometric observations), $R_{\rm BLR}$, yet $R_{\rm max} > R_{\rm BLR}$. Observing a large sample of Sy 1 galaxies with equatorial scattering \citet{APS19} found that $R_{\rm BLR}=(0.31\pm0.17)R_{\rm max}$. Also, $R_{\rm max}$ should be connected with the scattering region size $R_{\rm sc}$, and \citet{APS19} showed by observations that $R_{\rm sc}/R_{\rm max} = 1.72 \pm 0.48$. This relation is in a good agreement with the result found by simulations \citep{sa18}: $R_{\rm sc}/R_{\rm max} = 1.5 - 2.5$. Let us examine our result with these relations. Considering $R_{\rm BLR} \approx 20$  lt-days, one obtains $R_{\rm max} \approx 65$  lt-days. Then, the relation between the maximal BLR size and the size of the scattering region obtained by the suggested method is equal to $R_{\rm sc}/R_{\rm max} \approx 1.5$, which corresponds to the previous investigations. It also shows that the suggested method of the reverberation mapping in the polarized line gives the result within the frames of the earlier theoretical and practical works.}

{\textit{b) Comparison with $R_{\rm dust}$.} \citet{Kish11} estimated the inner size of the dusty region at 214$\pm$59  lt-days using observations at the Keck interferometer in the $K$ band. Also, \citet{Kish11} adduced an estimation of the dusty region size of $\sim$115  lt-days from UV luminosity using the fit by \cite{Suganuma06}. These introduced values differ $\sim$1.9 times but this difference is predicted from the common conceptions (see Fig. \ref{fig1}). The maximum of near-IR emission observed with $K$-band interferometry is coming from the dusty region parts that are further from the central source than the dust sublimation region, i.e. the region where the dust survives but does not dominate. The estimation of $R_{\rm sc}$ obtained within our work is pretty close to the value calculated from UV luminosity. It leads to two possibilities: (i) the variable activity of the central source varies the location of the equatorial scattering screen within the region between the BLR and the dusty region, therefore, the observations at different epochs give unbridgeable estimations; (ii) the equatorial scattering screen locates closer to the dusty region and the influence of the dust scattering is underestimated in modeling. }

\section{Conclusions}
\label{conc}

{Here we propose a new approach to measure the distance to the equatorial scattering region in Type 1 AGNs. Our method could be called a crucial "reload"\ of the one suggested by \citet{Gas12}. By subtracting the polarized signal from the polarized broad line flux, the contributions of the other polarization mechanisms, except the one due to the equatorial scattering, are minimized. So, the measurement of the time delay in the emission line in polarized light reveals the distance to the scattering region located between the BLR and the dusty region \citep{Smith04, GG07, sa18}. $R_{\rm sc}$ is a probing of the lower limit of the dust sublimation radius, and the attempt to locate the scattering region more precisely is an additional key to the understanding of the physical state inside AGNs.}

{The method has some advantages. 
\begin{itemize}
    \item  This is a method that directly measured $R_{\rm sc}$, and there are no more observational methods to directly determine, more or less, $R_{\rm sc}$.
    \item The suggested approach is the minimum affected by other polarization mechanism,s since broad lines seem to be equatorially scattered.
    \item The method seems to be inclination independent.
    \item As soon as the method is suitable for the broad emission lines up to Mg II \citep{unpublishedkeyA}, it can be used to probe the distances inside AGNs at the greater $z$ than the methods based on the IR observations as (i) the $K$-band interferometry is not able to resolve the central parts of far AGNs because of the limited angular resolution; (ii) going at greater redshifts, the regions observed in the $K$ band in local AGNs are shifted to the far-IR bands, and farther, which makes it harder to observe them with the same approach.
    \item The direct estimates of $R_{\rm sc}$ will increase the accuracy of mass estimations \citep{AfPo14}. 
\end{itemize}}

{However, the suggested approach has important limits:
\begin{itemize}
    \item The method could be used only for Type 1 AGNs with observable signatures of the equatorial scattering \citep[such objects could be found in, e.g.,][]{smith02, APS19}. 
    \item Even though the idea is simple, the influence of the other internal polarization mechanisms \citep[see][]{sa18} on the observed time lag should be carefully modeled.
    \item Also, the essential point is whether the activity of the ionizing source influences strongly on the $R_{\rm sc}$ and other component locations within the AGNs.
\end{itemize}}

{The offered approach would be further approved by observations of a sample of Type 1 AGNs. Also, the important point is to apply the photometric reverberation mapping technique \citep{haas11} to our method of reverberation mapping in polarized light, and to involve small-sized telescopes.}

\section*{Acknowledgements}
The reported study was funded by RFBR, project number 20-01-00001. The results of observations were obtained with the 6 m
BTA telescope of the Special Astrophysical Observatory of
Academy of Sciences, operating with the financial support of
the Ministry of Education and Science of Russian Federation
(state contracts no. 16.552.11.7028, 16.518.11.7073).  L. {\v {C}.}  Popovi\'c  is supported by the Ministry of Education, Science and Technological Development of R. Serbia (project No 176001). 

\bibliography{lit}



\end{document}